\documentclass{calor2002}

\begin{document}

\title{Borexino: A real time liquid scintillator detector for low energy solar
neutrino study}

\author{LINO MIRAMONTI}

\address{Physics Department of Milan University and INFN, \\
Via Celoria 16, 20133 Milano\\
E-mail: Lino.Miramonti@mi.infn.it}

\maketitle

\abstracts{
Borexino is a large unsegmented calorimeter featuring 300 tons of liquid
scintillator, contained in a 8.5 meter nylon vessel, viewed by 2200 PMTs.
The main goal of Borexino is the study, in real time, of low energy solar 
neutrinos, and in particular, the monoenergetic neutrinos coming from $^7Be$, 
which is one of the missing links on the solar neutrino problem.
The achievement of high radiopurity level, in the order of $10^{-16} g/g$ of U/Th 
equivalent, necessary to the detection of the low energy component of the 
solar neutrino flux, was proved in the Borexino prototype: the Counting 
Test Facility.
The detector is located underground in the Laboratori Nazionali del Gran 
Sasso in the center of Italy  at 3500 meter water equivalent depth.
In this paper the science and technology of Borexino are reviewed and its main
capabilities are presented.}

\section{Introduction}

Our Sun is essentially a self-confining nuclear fusion reactor whose production rate is regulated by 
the weak nuclear interaction. All the reactions can be summarized by a single one in which four 
protons combine into a $^4He$ nucleus: $4p \rightarrow \alpha + 2e^- + 2\nu_e$.
In this process two neutrinos are emitted and energy is released ($\simeq$ 26.6 MeV). 
About $6 \cdot 10^{10}$ neutrinos of solar origin hit a centimeter square of the Earth per second.

The discrepancy between the $\nu_{e}$ expectations of the solar neutrino flux, as calculated 
by the Solar Standard Model (SSM)\cite {bib:Bahcall}, and the experimental results, is known as the Solar 
Neutrino Problem (SNP). The recent results from helioseismology confirms the SSM to an accuracy 
within $1\%$.

A possible explanation of the SNP is that $\nu_{e}$ emitted from the fusion reactions oscillate 
while traveling from the Sun to the Earth and transmute in non electronic flavour neutrinos, with cross 
sections null or lower than those of electron neutrino. This oscillation is possible if neutrino has mass 
and mass eigenstate do not coincide with flavour eigenstate. 

The probability for an electron neutrino of energy E to conserve its flavor traveling a distance L 
is: $P = 1 - \sin^2{2\theta \cdot \sin{(L\Delta{m^2}/(4E))}}$, where $\theta$ is 
the mixing angle and $\Delta{m^2}$ the difference between the squared masses. 
This scenario (Vacuum Oscillation  scenario or VO), can explain the experimental data if the mixing 
angle is such that $sin^2{2\theta} \simeq 1$ and the $\Delta{m^2} \simeq 10^{-11} eV^2$.
In case of resonant oscillation in the core of the Sun (Mikeyev-Smirnov-Wolfenstein effect or MSW effect), 
two different scenario are obtained; Large Mixing Angle solution (LMA) for $sin^2{2\theta} \simeq 1$ 
and $\Delta{m^2} \simeq 10^{-6} eV^2$, or Small Mixing Angle solution (SMA) for 
$sin^2{2\theta} \simeq {10^{-2}}$ and $\Delta{m^2} \simeq 10^{-6} eV^2$.

\section{The Borexino detector}

\subsection{The Borexino design}

\begin{figure}[th]
\centerline{\epsfxsize=2.9in\epsfbox{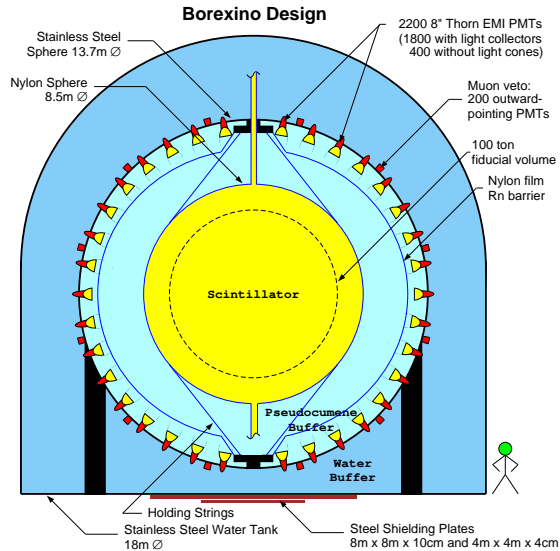}}   
\caption{Schematic view of Borexino. \label{fig:borex_scketch}}
\end{figure}

The construction philosophy adopted by our collaboration is that of a graded shield of 
progressively lower radioactivity material approaching the detector's core, ending in the 
definition of a fiducial volume.
The apparatus is located deep underground at about 1780 m of overburden rock (about 3500 meter 
water equivalent) in the Apennines in the Laboratori Nazionali del Gran Sasso of the 
National Institute of Nuclear Physics (INFN) in the center of Italy. The muons crossing 
the detector is about 1.1 per $m^2$ $h^{-1}$ corresponding to a reduction of about six 
order of magnitude.
The core of the detector consists in 300 tons of liquid scintillator contained in a 
125 $\mu m$ transparent nylon vessel (Inner Vessel) of 8.5 m diameter, viewed 
by 2200 photomultiplier tubes supported on a stainless steel sphere (SSS) of 13.7 m diameter. 
This sphere is than enclosed in a cylindrical tank 17 m height with a diameter of 18 m. 
The sketch of the apparatus is presented in Figure~\ref{fig:borex_scketch}.

The zone between the Inner Vessel and the photomultiplier tubes is filled with pure 
pseudocumene (with a 5 g/l shift wavelenghter) in order to compensate the buoyancy 
force and  to ensure good light coupling; furthermore in pseudocumene the radiopurity 
achievable is greater than in water.
The region between the stainless steel sphere and the external tank is filled with ultrapure 
water in order to reduce neutrons and gammas coming from the surrounding rock. 
Furthermore the ultrapure water contained in the external tank serves also for muon veto.
Between the stainless steel sphere and the Inner Vessel a second nylon vessel (Outer Vessel) 
is planned in order to stop the radon emanating from the outer part of the detector and in 
particular from the photomultiplier tubes.
There are three different classes of photomultiplier tubes; the first one (1800 inner PMTs) 
is composed by tubes equipped with a light concentrator designed 
in such a way to observe only photons coming from the Inner Vessel. A second class is 
composed by 400 PMTs without light concentrator, that in addition to the light from the 
Inner Vessel are able to detect the light produced by events in the pseudocumene buffer. 
The task of these inner PMTs is that of survey for the light produced by muons that could 
be confused with neutrino signal.
The last class of photomultiplier tubes is composed by 200 outer PMTs placed on the 
outer surface of the stainless steel sphere. The goal of these devices is to identify muons 
both traversing and not traversing the pseudocumene buffer. In the first case, they allow 
a double tag to identify such events, and for the second case they give an alert to search 
afterward for muon induced events which could reach the scintillator.

\subsection{The photomultiplier tubes}
The photomultiplier tubes employed in Borexino are 8 inchs 9351 made by ETL (former EMI). 
The main characteristics are summarized in Table~\ref{tab:PMT}. Among its performances we can notice 
a limited transit time spread (jitter) that is less than 1 ns, a good peak to valley ratio (2.5) a 
very low dark noise rate (1 kHz), a low after-pulsing probability ($\leq 4 \%$). 
Furthermore, the strong requirements about radiopurity led us to manufacture these PMTs with 
special low radioactivity glass and internal parts\cite{bib:RanucciPMT}.
The 1800 light concentrators are realized with anodized aluminium in order to have a good 
reflectivity in the 400 nm region and a good compatibility with pseudocumene.

\begin{table}[h]
\tbl{Characteristics of Borexino 8 inch photomultiplier tubes.\vspace*{1pt}}
{\footnotesize
\begin{tabular}{|l|r|}
\hline
{} &{}  \\[-1.5ex]
{} & $Characteristics$\\[1ex]
\hline
{} &{}  \\[-1.5ex]
Quantum efficiency &$26\%$ at 420 nm \\[1ex]
Transient time spread &1 ns            \\[1ex]
Peak/valley &2.5              \\[1ex]
Dark noise &1 kHz            \\[1ex]
Gain &$10^7$             \\[1ex]
Low radioactivity glass &schott 8246 \\[1ex]
\hline
\end{tabular}\label{tab:PMT} }
\end{table}

\subsection{The liquid scintillator}
\begin{figure}[!ht]
    \begin{center}
        \rotatebox{270}{\includegraphics [width=0.60\textwidth,
        clip=true,bb=95 2 600 690]{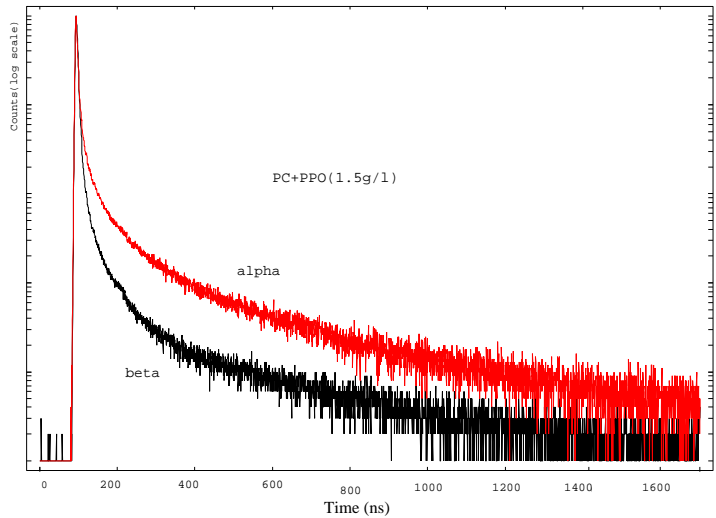}}
    \caption[Dis]{Experimental light profiles measured for the scintillator (PC with PPO at 1.5 g/l for
alpha and beta particles.\label{fig:sci_decay}}
    \end{center}
    \end{figure}

The liquid scintillator is composed by Pseudocumene as solvent and PPO at the concentration 
of 1.5 g/l as solute. The pseudocumene (PC) is 1,2,4-trimethylbenzene $C_6H_3(CH_3)_3$, and the PPO
is 2,5-diphenyloxazole $C_{15}H_{11}NO$. The refractive index of the scintillator is equal to 1.505.
The excitations, produced by electrons ($\beta$)/gamma rays ($\gamma$) and alpha particles ($\alpha$), have
different properties (see Table~\ref{tab:LiquidScint}).
During the research and development period a great numbers of tests were performed in order 
to study the intrinsic optical properties of the scintillator\cite{bib:Elisei}. At the concentration of 1.5 g/l, the light 
yield results to be about 11000 photons per MeV,  with a fast decay time of about 3.6 ns (the intrinsic 
decay time of the PPO is 1.6 ns, and those of pseudocumene without solute is about 28 ns). 
The attenuation length is about 30 m and the scattering length is about 7 m, both for a wavelength of 420 nm.
This mixture gives also a good alpha/beta discrimination (see Figure~\ref{fig:sci_decay}).

\begin{table}[h]
\tbl{The decay times and the relative quantities of the decay components
of the liquid scintillator PC+PPO at 1.5 g/l.\vspace*{1pt}}
{\footnotesize
\begin{tabular}{|l|r|r|r|r|}
\hline
{} &{} &{}&{} &{}\\[-1.5ex]
{}Component  & $\beta$ and $\gamma$ & &$\alpha$ & \\[1ex]
\hline
\hline
{} &{} &{} &{} &{}\\[-1.5ex]
$First$  &3.574 ns &89.52 $\%$  &3.254 ns &63.00 $\%$ \\[1ex]
$Second$ &17.61 ns &6.33  $\%$  &13.49 ns &17.82 $\%$ \\[1ex]
$Third$  &59.49 ns &2.93  $\%$  &59.95 ns &11.93 $\%$ \\[1ex]
$Fourth$ &330.4 ns &1.22  $\%$  &279.1 ns &7.25  $\%$ \\[1ex] 
\hline
\end{tabular}\label{tab:LiquidScint} }
\end{table}

\subsection{The Borexino detector performance}\label{sec:plac}
The PMT's total coverage is about $34\%$, hence, taking into account the photocathode 
efficiency, the photoelectron yield is about 450 pe/MeV (at 1 MeV), giving an 
energy resolution of $8\%$ (at 1 MeV) and a spatial resolution of 12 cm 
(at 1 MeV). These good performances, high energy and spatial resolution are 
very important for the experiment; the energy resolution is essential to 
disentangle the recoil spectrum of the scattered electrons from the incoming 
neutrinos via the sharp edge at about 660 keV. The high spatial resolution 
permits a good spatial reconstruction of the events and hence to reject the 
background coming from outside.

\subsection{Calibrations and monitoring of the detector}
In order to assure a careful understanding of the detector operational conditions during the data taking, 
different calibrations and monitoring systems are envisaged\cite{bib:calib}.
For a precise event reconstruction, it is needed a precise time measurement of the PMTs signals. 
We plan to perform this measurement by illuminating each photomultiplier tube with an optical fiber 
coupled to a laser generating a fast pulse of the order of 50 ps.
It will be also monitor the gain stability of the PMTs since these last are illuminated at the level of single 
photoelectron.
In order to monitor in time the stability of the scintillator, we plan to locate an external $^{228}Th$ source 
just outside the stainless steel sphere.  
The scintillator will be monitored also via radon sources inserted inside the inner vessel through organ pipes.
In addition, a continuous check of the stability of the optical properties of the buffer is performed by means 
of laser light that induce photoexcitation process.

\section{The physics of Borexino}
The main goal of Borexino is the detection of solar neutrinos coming from the electron capture of $^7Be$  
that gives a monochromatic line at 862 keV through the $ \nu + e^- \rightarrow \nu + e^- $ electroweak scattering reaction.
The high radiopurity level and the high yield of light of the scintillator make it possible to reach a 
detection threshold as low as 250 keV\footnote{The 250 keV energy threshold is dictated by the
$^{14}C$ content in the organic scintillator. The $^{14}C/^{12}C$ must be in the range of $10^{-18}.$}. 
The scattered electron produces a continuum recoil spectrum, with a maximum energy of about 660 keV.
The expected event rate, according to the Solar Standard Model in the case of non-oscillation, is about 
55 events per day in 100 tons of fiducial volume.
The ultimate background component is represented by the scintillator itself; the U and Th content must 
be kept at the level of $10^{-16} g/g$, while for the natural K at the level of $10^{-14} g/g$. 
This level of impurities will give, in 100 tons fiducial volume, about 10 events per day.

\begin{figure}[th]
\centerline{\epsfxsize=2.2in\epsfbox{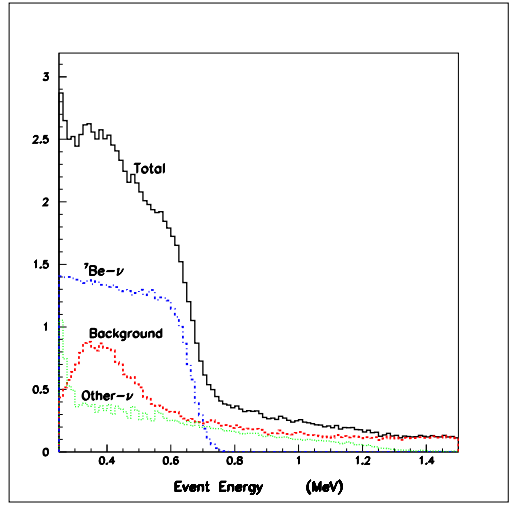}}
\caption{The estimate background spectrum and the Solar Standard Model neutrino signal. \label{fig:ctotalrate}}
\end{figure}

Figure~\ref{fig:ctotalrate} shows the estimate background spectrum and the Solar Standard Model neutrino signal.

In Table~\ref{tab:neutrinorates} the solar neutrino counting rates expected per day 
in the Borexino neutrino window (i.e. 250 keV - 800 keV) for 4 scenario are reported: The Solar Standard Model 
(SSM), the Large Mixing Angle solution (LMA), the Small Mixing Angle solution (SMA) and the low mass solution (LOW).

\begin{table}[h]
\tbl{Solar neutrino counting rate per day in Borexino fiducial volume in the neutrino energy window (250 keV - 800 keV)
 for 4 scenarios: The Solar Standard Model (SSM), the Large Mixing Angle solution (LMA), for the Small Mixing 
Angle solution (SMA) and for the low mass solution (LOW). These rates are calculated for a mean Sun-Earth 
distance and 113 $m^3$ fiducial volume. The  $\Delta{m^2}$ and $sin^2{2\theta}$ used parameters are:
   $\Delta{m^2} \simeq 1.8 \cdot 10^{-5} eV^2$ and $sin^2{2\theta}=0.76$  for LMA, 
  $\Delta{m^2} \simeq 5.4 \cdot 10^{-6} eV^2$ and  $sin^2{2\theta} \sim 5.5 \cdot {10^{-3}}$ for SMA and 
  $\Delta{m^2} \simeq 7.9 \cdot 10^{-8} eV^2$ and $sin^2{2\theta}=0.96$
   for LOW.\vspace*{1pt}}
{\footnotesize
%\tabcolsep7pt
%\begin{tabular}{@{}crrrr@{}}
\begin{tabular}{|c|r|r|r|r|r|r|r|r| }
\hline
{} &{} &{} &{} &{} &{} &{} &{} &{} \\[-1.5ex]
{} & $pp$ & $^7Be$ & $pep$ & $^{13}N$ & $^{15}O$ & $^{17}F$ & $^8B$ & $total$ \\[1ex]
\hline
\hline
{} &{} &{} &{} &{} &{} &{} &{} &{}\\[-1.5ex]
$SSM$ &0.22 &43.3 &2.0  &4.0  &5.5  &0.07 &0.08 &55.2\\[1ex]
$LMA$ &0.15 &24.4 &0.95 &2.27 &2.86 &0.03 &0.03 &30.7\\[1ex]
$SMA$ &0.08 &9.2  &0.39 &0.87 &1.12 &0.01 &0.04 &11.7\\[1ex]
$LOW$ &0.13 &22.8 &1.03 &2.13 &2.86 &0.03 &0.04 &29.0\\[1ex] 
\hline
\end{tabular}\label{tab:neutrinorates} }
\end{table}

\section{The Counting Test Facility}
In order to prove the capability to reach a so high radiopurity level, a prototype of the 
detector was constructed: the Counting Test Facility (CTF).
The main goal of CTF was to develop solutions directly applicable to operational issues 
for Borexino(for further details see\cite{bib:CTF1}).

In CTF 4 tons of liquid scintillator are contained in a 2 m diameter transparent nylon vessel 
mounted at the center of an open structure that supports 100 PMTs. These last are 
coupled to optical concentrators viewing the nylon vessel with $20\%$ optical 
coverage. The whole system is placed within a cylindrical tank (10 m of height and 
11 m of diameter) that contains about 1000 tons of ultra-pure water in order to shield 
against gamma rays coming from the PMTs and other detector materials and neutron from the rock.
With 300 pe/MeV the CTF reachs an energy resolution of $9\%$ at 825 keV 
($^{214}Po$ line) and a spatial resolution of about 12 cm at the same energy. 
Concerning the $\alpha/\beta$ discrimination capabilities, the $\alpha$ identification 
efficiency is $95\%$ and the $\beta$ inefficiency is of few percents.
The radiopurity levels evaluated in CTF are:

\begin{center} 
$ ^{238}U \leq 3.5 \pm 1.3 \cdot 10^{-16} g/g $ \\
 
$ ^{232}Th = 4.4^{+1.5}_{-1.2} \cdot 10^{-16} g/g $ \\
 
$ ^{14}C/^{12}C = 1.94 \pm 0.09 \cdot 10^{-18}.$ \\
\end{center}

\section{Conclusions}
The demands on the radiopurity of the detector materials, especially for the scintillator itself, are very 
challenging. The CTF has demonstrated the capability to achieve unprecedented levels of radiopurity 
in the liquid scintillator, opening in this way the construction of Borexino. Today, Borexino is in its 
construction final phase, and probably it will be able to take physical data from the spring of the next year 
playing in this way a crucial role in the understanding of the solar neutrino puzzle.

\section*{Acknowledgments}
The author acknowledges the support of the Italian National Institute of Nuclear Physics.

\section*{Collaboration list}
Belgium: I.R.M.M. European Joint Research Center - Geel. Canada: Queen's University - Kingston. 
France:  Coll\'{e}ge de France. Germany: Max-Planck-Institut fuer Kernphysik Heidelberg, 
Technische Universitaet Muenchen. Hungary: KFKI-RMKI Research Institute for Particle and Nuclear Physics Budapest. 
Italy: Dipartimento di Fisica Universit\'{a} and I.N.F.N di Genova, L.N.G.S., Dipartimento di Fisica Universit\'{a} 
and I.N.F.N. di Milano, Dipartimento di Fisica Universit\'{a} and I.N.F.N. di Pavia, Dipartimento di Chimica 
dell'Universit\'{a} and I.N.F.N di Perugia. Poland: Institute of Physics, Jagellonian University Cracow. 
Russia: J.I.N.R. DUBNA, Kurchatov Institute - Moscow. United States: Bell Laboratories Lucent Technologies, 
Massachusetts Institute of Technology, Princeton University, Virginia Polytechnic Institute.


\begin{thebibliography}{0}

\bibitem{bib:Bahcall} J.N. Bahcall et al., Astrophys J.555:990-1012, (2001) and references therein.

\bibitem{bib:RanucciPMT} G. Ranucci et al., Nucl. Inst. and Meth. A 333 (1993) 553-559.

\bibitem{bib:Elisei} F. Elisei et al., Nucl. Inst. and Meth. A 400 (1997) 53-68.

\bibitem{bib:CTF1} G. Alimonti et al., Nucl. Inst. and Meth. A 406 (1998) 411-426.

\bibitem{bib:calib} L. Miramonti, Progress in Part. and Nucl. Phys. 48 (2002) 27-28.


\end{thebibliography}
\end{document}